# Temporo-Spatial Collaborative Filtering for Parameter Estimation in Noisy DCE-MRI Sequences: Application to Breast Cancer Chemotherapy Response

Xia Zhu[a*], Dipanjan Sengupta[a], Andrew Beers[b], Kalpathy-Cramer Jayashree[bc], Theodore L. Willke[a]

[a]Intel Corporation, New Technology Group, Hillsboro, OR, USA
[b]Athinoula A. Martinos Center for Biomedical Imaging, Department of Radiology, Massachusetts General Hospital and Harvard Medical School, Boston, MA, USA
[c]The Center for Clinical Data Science, Massachusetts General Hospital and Brigham and Women's Hospital, Boston, MA, USA

**Abstract.** Dynamic contrast-enhanced magnetic resonance imaging (DCE-MRI) is a minimally invasive imaging technique which can be used for characterizing tumor biology and tumor response to radiotherapy. Pharmacokinetic (PK) estimation is widely used for DCE-MRI data analysis to extract quantitative parameters relating to microvasculature characteristics of the cancerous tissues. Unavoidable noise corruption during DCE-MRI data acquisition has a large effect on the accuracy of PK estimation. In this paper, we propose a general denoising paradigm called *gather-noise attenuation and reduce* (GNR) and a novel temporal-spatial collaborative filtering (TSCF) denoising technique for DCE-MRI data. TSCF takes advantage of temporal correlation in DCE-MRI, as well as anatomical spatial similarity to collaboratively filter noisy DCE-MRI data. The proposed TSCF denoising algorithm decreases the PK parameter normalized estimation error by 57% and improves the structural similarity of PK parameter estimation by 86% compared to baseline without denoising, while being an order of magnitude faster than state-of-the-art denoising methods. TSCF improves the univariate linear regression (ULR) c-statistic value for early prediction of pathologic response up to 18%, and shows complete separation of pathologic complete response (pCR) and non-pCR groups on a challenge dataset.

*Xia Zhu, xia.zhu@intel.com

## 1 Introduction

Dynamic contrast-enhanced magnetic resonance imaging (DCE-MRI) is a well established method for measuring changes in microvascular properties associated with diseased tissues in order to characterize tissue biology.[1–15] Typically, a bolus of a diffusible contrast agent (CA) is injected and the physiological information is estimated by the pharmacokinetic (PK) model, usually a compartmental model describing the rate of transfer of contrast agent between blood pool and extravascular extracellular space (EES). With the modern fast sequences DCE-MRI can provide very good temporal and spatial resolution along with good organ coverage[6] and hence DCE-MRI becomes a promising minimally invasive imaging technique to characterize tumor biology and tumor response to radiotherapy.[16] The recommended kinetic parameters of interest[17–19] in anti-angiogenic and anti-vascular therapies are usually $K^{trans}$, the transfer constant of the CA, and $v_e$, the volume fraction of EES.

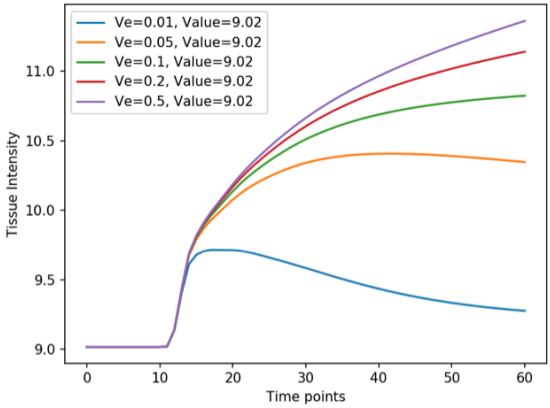 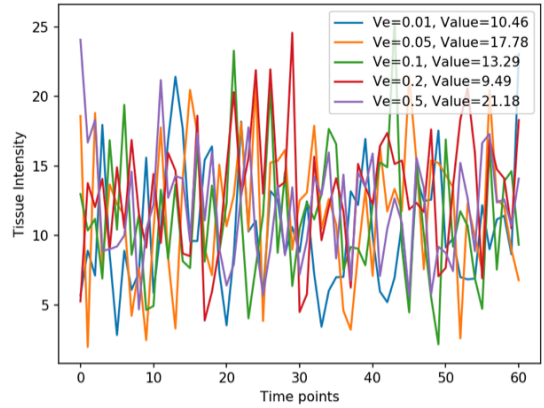

(a) DRO v9 ground truth signals.        (b) DRO v9 noisy signals.

Fig 1: DCE-MRI signals with and without noise ($K^{trans} = 0.01$). Values shown in the figure are pre-contrast signal intensity. Data is digital reference object (DRO) data from Duke University.

Adoption of DCE-MRI into widespread clinical use requires accurate estimation and good reproducibility of the pharmacokinetic parameters. Although the parameters estimated from pharmacokinetic modeling of DCE-MRI data sequence should in theory be independent of data acquisition platforms and techniques, the accurate estimation of these parameters is influenced by many factors, such as choice of arterial input functions (AIF or the concentration of CA in the blood pool as a function of time),[20–25] quantification of pre-contrast relaxation rate $R_1$,[21,26–29] selection of model and techniques for precise parameter estimation using the selected model to fit the data,[18,30–33] and the signal-to-noise ratio (SNR).[28,34–41]

This paper focuses on the inaccuracies in the estimation of kinectic parameters due to unavoidable noise corruption at the DCE-MRI data acquisition step and proposes methods and algorithms to handle such scenarios. The MRI raw data is measured through a quadrature detector resulting in the real and the imaginary quadrature channels each separately corrupted with white additive noise. Taking the magnitude of the complex data (Rician distribution) results in a nonlinear mapping between the noisy and the true unknown signal rendering the noise non-additive. Hence one of the key challenges in denoising MR data corrupted with Rician noise is heteroskedasticity (i.e., the standard-deviation of the noise corrupted magnitude signal is not uniform across the entire image but depends on the unknown true signal itself). As the PK parameter estimation is directly dependent on the signal intensity (Pharmacokinetic and MRI model, appendix), the noise distorted signal has direct implications on the accuracy of the estimated parameters, and thereby the characterization of the tissue biology. Figure 1 shows how MRI acquisition noise can adversely affect the intensity of DCE-MRI data, and thus affect the accuracy of PK parameters estimation.

To address the above mentioned challenges in noise corrupted MRI data, several denoising schemes have previously been proposed to estimate the true signal from the raw magnitude MRI signal. Many of them are maximum-likelihood estimation (MLE) based methods for simultaneous noise variance and signal intensity estimation from MR magnitude imaging.[36–39] However, because such approaches require nonlinear optimization of a 2D function at each voxel, they come with a high computational cost. Another approach involves calculation of an unbiased estimator of the signal intensity that depends on the second moment of the magnitude signal.[40,41] However, this works on the assumption that the noise variance is known a priori, which might not be realistic in a real world scenario. Koay et al.[34] proposed a fixed point formulation of signal-to-noise ratio (SNR) and a correction factor that establishes a relationship between the magnitude MR variance and the variance of the underlying Gaussian noise in the quadrature channels. Hence the signal intensity and the noise variance can be extracted simultaneously from a noisy magnitude MR signal. Although this method is computationally very efficient, it only works for homogeneous region of interest (ROI) signals, and the fix point formula isn't valid for all ranges of SNR and hence isn't always practical to deploy.

Several bodies of work on denoising have been proposed where the noise is assumed to be additive white noise. BM3D[42] is one such method that groups spatially similar 2-D image fragments into 3-D groups and collaboratively filters these 3-D groups. Although many algorithms have been proposed since BM3D, with a few claiming better performance than BM3D, recent research shows that many that perform well on synthetic noise are outperformed by BM3D on images with real noise.[43] Therefore, BM3D arguably remains the state-of-the-art image denoising method for white noise. BM4D[44] is an extension of BM3D filter to volumetric data where mutually similar 3-D patches are stacked together in a 4-D array and jointly filtered in a transform domain. While in BM3D the basic data patches are blocks of voxels, BM4D utilizes cubes of voxels, which are stacked into a 4-D group. The key drawbacks of BM3D and BM4D filtering are two-fold: first, they work poorly in scenarios when directly applied to non-additive noise corrupted images like in the case of MR images. Second, although they make use of spatial similarity information within an image, they ignore temporal correlation across multiple inter-dependent image sequences (e.g. scans in DCE-MRI data).

In this paper we propose *Temporo-Spatial Collaborative Filtering* (TSCF), a novel denoising technique that enhances sparse representation based on both temporal correlation and spatial similarity of DCE-MRI time series. The sparsity is achieved by *gathering* similar voxels across the entire time-series into clusters, hence preserving both the spatial and temporal similarity information in them. These clusters are then collaboratively filtered for *noise attenuation*. This involves three sub-steps: cluster transformation, noise attenuation and inverse transformation outputting a jointly filtered denoised set of clusters. By taking into account both temporal and spatial similarity in the voxels grouped in a cluster, our denoising method achieves enhanced sparsity in the transform domain and the noise can be attenuated by preserving a few high-valued transform coefficients. Finally, we unravel the clusters to return each voxel to their original positions in the image. Because there exists a many-to-one mapping between clusters and voxel positions in the image (i.e., each voxel can belong to multiple clusters), we perform a *reduce* operation on intensity contributions from multiple clusters to get the final estimation of the denoised signal for each voxel. We propose *gather-noise attenuation and reduce* (GNR) as a general denoising paradigm. TSCF is an instantiation of GNR for DCE-MRI data.

In order to address the challenges associated with heteroskedasticity of DCE-MRI data and to apply our proposed TSCF algorithm that works in the homoskedastic domain, we transform the raw data using variance stabilization transformation (VST).[35] After VST, the signal-dependent noise in DCE-MRI can be treated as additive with unitary variance which ensures the sequential noise estimation and denoising using TSCF. Finally the inverse-VST is applied to recover the denoised signal, followed by PK parameter estimation to achieve optimal results.

To establish the efficacy of our proposed method and reproducibility of its results we designed our experiments as per the guidelines of Quantitative Imaging Network (QIN)[31,45] that the National Cancer Institute has recently established for development and validation of quantitative imaging tools. Experimental results of DCE-MRI data analysis, performed on a dataset from a digital reference object (DRO)[45] and human breast tumor dataset, demonstrate that our TSCF denoising algorithm decreases the PK parameter normalized estimation error by 57% and improves the structural similarity of PK parameter estimation by 86% compared to baseline without denoising. Our method beats the state-of-the-art denoising technique, BM4D, by taking into account both temporal correlation and spatial similarity information which was missing in previous DCE-MRI denoising strategies. Besides accuracy improvement, our TSCF denoising is an order of magnitude faster than BM4D. Finally, TSCF improves the univariate linear regression (ULR) c-statistic value for early prediction of pathologic response up to 18%, and shows complete separation of pathologic complete response (pCR) and non-pCR groups on breast cancer dataset.

To summarize our main contributions:

- We propose a 3-stage end-to-end denoising and PK parameter estimation pipeline for noise corrupted DCE-MRI data.

- We proposed a general paradigm for signal restoration: gather-noise attenuate and reduce (GNR). As an instantiation of GNR, we also propose *Temporo-Spatial Collaborative Filter* (TSCF) denoising algorithm for noisy DCE-MRI, which achieves up to a 57% reduction in normalized estimation error, and up to 86% improvement in the structural similarity of PK parameter estimation than baseline method, and an order of magnitude speed up in execution time than state-of-the-art denoising methods.

- We evaluate the proposed framework on both synthetic DRO phantom data and real human breast data.

- We analyzed the early prediction of therapy response with and without denoising. TSCF improves the ULR c-statistic value up to 18% for early prediction of pathologic response, and is able to completely separate pCR and non-pCR groups.

The remainder of the paper is organized as follows: Section 2 discusses the proposed Temporo-Spatial Collaborative Filtering (TSCF) technique for noise removal and pharamacokinectic param-eter estimation. Section 3 presents the experimental setup and results. Section 4 discusses the key takeaways from the experimental results followed by conclusions and future work.

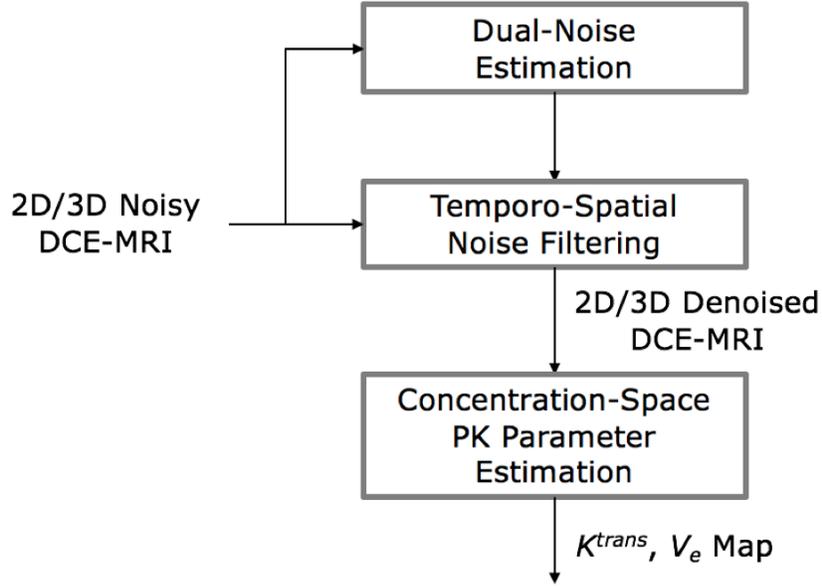

Fig 2: Pipeline for Noisy DCE-MRI Data

## 2 Proposed Methods

As shown in Figure 2, our proposed DCE-MRI parameter estimation pipeline consists of three stages. In the first stage, we estimate the statistics of noise corrupted DCE-MRI data using a *dual-noise estimation* technique assuming that the raw input data belongs to the Rician distribution.[46] In the second stage, our novel *temporo-spatial collaborative filtering* technique accurately and efficiently denoises the DCE-MRI raw data and restores the original signal by making use of both temporal correlation and spatial similarity between pixels (or voxels; note that our algorithm works for both 2-D and 3-D medical images. We use voxels in the following text for simplicity). Finally, using the denoised data from the previous stage, Pharmacokinetic (PK) parameters are estimated in a very computationally efficient manner by working in the *concentration domain* instead of intensity domain. Next we discuss each of these pipeline stages in detail.

## 2.1 Dual Noise Statistics Estimation

In the first stage of our pipeline, we estimate the noise statistics of the input data using a dual noise estimation strategy, which is not only accurate but also computationally efficient by combining the best of two existing methods. As shown in Appendix Equation (29), the magnitude of MR images can be modeled by the Rice distribution,[47] which has two parameters: unknown true signal ($y$), and standard deviation of the additive noise ($\sigma_g$) that corrupts the real and imaginary part of the input data.

Koay et al.[34] proposed an analytically exact fixed point formulation of signal-to-noise ratio (SNR), $\theta$, and a correction factor $\xi(\theta)$ (Equation (32), appendix) that fundamentally relates the variance of the magnitude MR signal ($\sigma_m^2$) and the variance of the underlying Gaussian noise in the two quadrature channels ($\sigma_g^2$). Therefore, given the mean of the magnitude MR signal, $\mu_m$, and $\sigma_m$, using equation (32), one can determine the value of SNR in the image. We employ Newton's method of root finding, as described in [34] to estimate the SNR, and thereby calculate $y$ and $\sigma_g$. Although, the above mentioned method is computationally very efficient for calculating the SNR, it has two major limitations. First, the fixed point formula has a valid unique solution only when $\frac{\mu_m}{\sigma_m} \geq \sqrt{\frac{\pi}{(4-\pi)}} = 1.91$. Therefore, in scenarios when $\frac{\mu_m}{\sigma_m}$ is less than this threshold, the above method will fail to estimate $\sigma_g$. Second, the first moment of noisy data is an ensemble average over the noise fluctuations. Therefore this method works on calculating statistics of repeated measurement of the same noiseless signal, or a ROI-based averaging when there are homogeneous signals within the ROI.[34] The limitations bring us to another noise-level estimation technique based on variance-stabilization.

Foi et al.[35] proposes variance stabilization transform (VST) that maps heteroskedastic MR data to a homoskedastic domain assuming that standard deviation of the noise $\sigma_g$ is known. Further, Foi[35] employs a recursive approach to estimate the noise-level by iteratively updating $\sigma_g$ so that the standard deviation of the transformed data approaches some constant (e.g. 1). As each $\sigma_g$ update iteration involves a VST computation, this noise-level estimation strategy requires longer execution time than Koay's[34] closed-form solution.

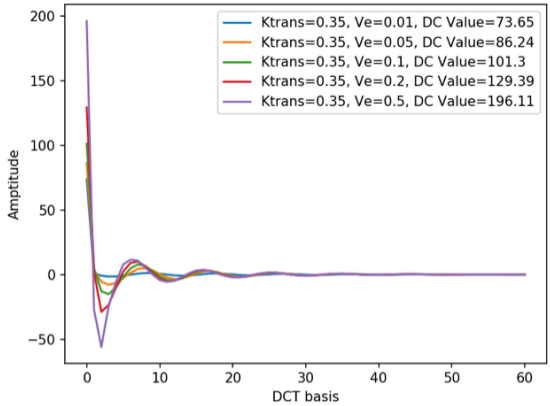 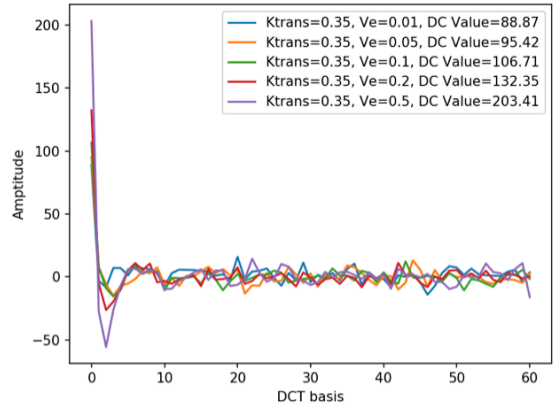

(a) DCT of DRO v9 ground truth signals.  (b) DCT of DRO v9 noisy signals.

Fig 3: Temporal correlation of DCE-MRI time series leads to energy concentration at low frequencies.

In our dual noise estimation stage, if there are homogeneous signals within the ROI, first we calculate the ratio $\frac{\mu_m}{\sigma_m}$ from the noise corrupted raw magnitude signal. If this is greater than a given threshold (1.91), we estimate the SNR ($\theta \equiv \frac{y}{\sigma_g}$), using a fixed point formula (Equation (32), appendix). Else we employ the computationally more intensive VST-based method to iteratively estimate the standard deviation of the noise.

To summarize, in dual-noise estimation we make full use of the benefits of two existing methods. We first try the fixed point formula if the signal within the ROI is homogeneous (based on prior or domain knowledge) and the magnitude SNR is in the valid range. Otherwise, we employ the computationally more intensive VST-based iterative solution.

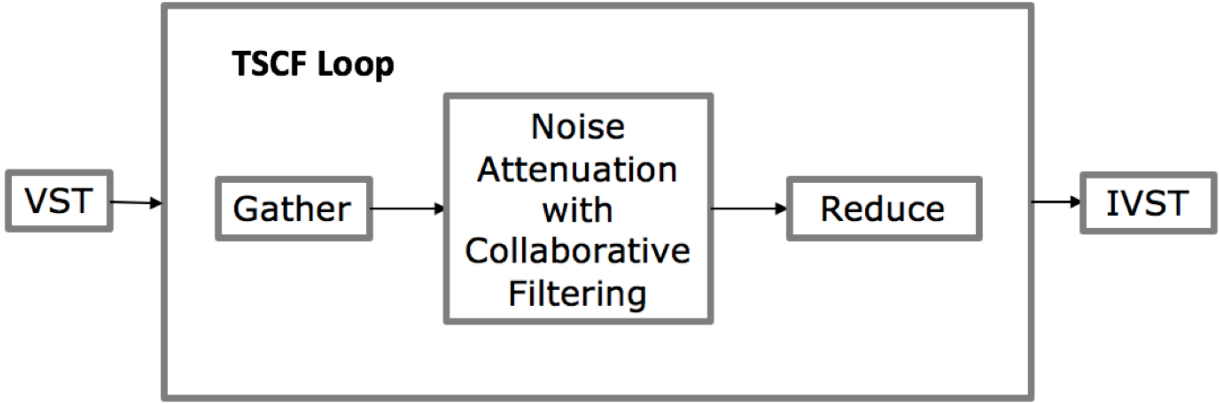

Fig 4: TSCF Pipeline.

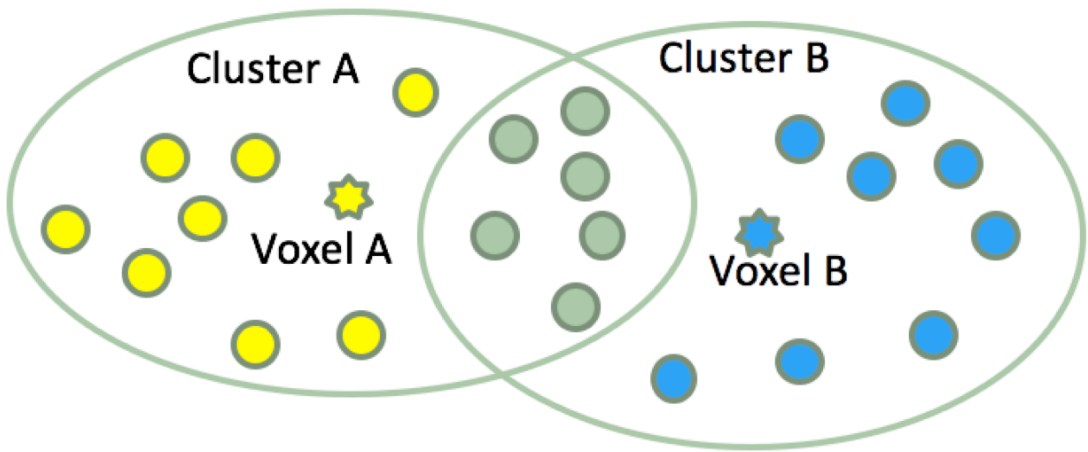

Fig 5: TSCF Gather.

## 2.2 Signal Restoration

The second stage of our pipeline is responsible for denoising the DCE-MRI input data using the noise statistics estimated by the previous stage. Here we introduce a more general denoising paradigm, *gather-noise attenuation and reduce* (GNR), using which a class of algorithms for signal restoration can be implemented. GNR requires the algorithm designer to define three basic primitives: 1) *gather* operation in which pixels (or voxels) are grouped together using an algorithm specific similarity or affinity metric. 2) *noise attenuate* operation is a group operation that uses special properties of these grouped pixels e.g. sparsity in the spectrum domain to collaboratively de-noise them. Finally, 3) *reduce* operation is responsible for recovering true pixel intensities as some algorithm specific function of the contributions from different denoised groups. The proposed temporo-spatial collaborative filtering (TSCF) algorithm is an instantiation of the GNR paradigm, specifically tailored to efficiently handle the task of denoising the DCE-MRI data sequence which has both temporal and spatial information incorporated in them.

The key idea behind our proposed TSCF algorithm is to extract sparsity from the temporal correlation and spatial similarity among the voxels in the DCE-MRI data. The signal restoration stage can be broadly divided into three sub-steps: 1) A preprocessing step to apply VST[35] to remove the dependency of the noisy signal variance on the true underlying signal. In other words, it transforms Rician noisy data to a signal with additive white noise. 2) Temporo-Spatial collaborative filtering to remove the white noise and 3) finally a post-processing step to apply unbiased inverse VST (IVST) on the denoised signal, to obtain the estimate of the signal of interest.

*2.2.1 Temporal correlation in DCE-MRI time series*

After applying VST, the noisy signal model for the input image of TSCF is

$$z(x,t) = y(x,t) + \eta(x,t), \quad x \in \mathbf{X} \tag{1}$$

where $z$ is the variance stabilized noisy image, $y$ is the variance stabilized noiseless signal and $x$ is a 2-D or 3-D coordinate belonging to the signal domain $\mathbf{X} \subset \mathbb{Z}^d$ and $d \in \{2,3\}$. $t \in \{0,1,...,N\}$ is the scan index, with $N$ as the total number of scans, and $\eta \sim \mathcal{N}(0, \sigma_v)$ is Gaussian white additive noise with zero-mean and variance $\sigma_v$.

Based on Tofts model,[2] from a spoilt gradient echo sequence, the intensity of DCE-MRI signal due to CA injection is modeled by Equation (2)

$$y(x,t) = y_0(x)\frac{\left(1 - e^{-TR/T_1}\right)\sin\theta}{1 - e^{-TR/T_1}\cos\theta} \qquad (2)$$

$$y(x,t) = y_0(x)\frac{\left(1 - e^{-TR/T_{10}}e^{-TR*r_1*C_t(x,t)}\right)\sin\theta}{1 - e^{-TR/T_{10}}e^{-TR*r_1*C_t(x,t)}\cos\theta} \qquad (3)$$

where $y_0(x)$ is the relaxed signal, $TR$ is the repetition time, $\theta$ is the flip angle, $C_t$ is the concentration of contrast agent in the tissue, $T_1$ is longitudinal relaxation time of the tissue, $T_{10}$ is longitudinal relaxation time before injection of contrast agent, and $r_1$ is the relaxivity.

Let

$$A = e^{-TR/T_{10}} \qquad (4)$$

$$B = A\cos\theta \qquad (5)$$

$$D = -TR*r_1 \qquad (6)$$

so that

$$\frac{y(x,t)}{y(x,t-1)} = \frac{\left[1 - Ae^{D*C_t(x,t)}\right]\left[1 - Be^{D*C_t(x,t-1)}\right]}{\left[1 - Be^{D*C_t(x,t)}\right]\left[1 - Ae^{D*C_t(x,t-1)}\right]} \qquad (7)$$

Equation (7) clearly shows that DCE-MRI time series of each voxel is temporally correlated. Furthermore, $C_t(x,t)$ and $C_t(x,t-1)$ are also correlated as shown in following equations.

$$K_{trans'}(x) = K^{trans}(x) \triangle t \tag{8}$$

$$E(x) = e^{-K_{trans'}(x)/V_e(x)} \tag{9}$$

$$F(x,t) = C_p(t-1)\left[E(x) - E(x)logE(x) - 1\right]$$

$$+ C_p(t)\left[E(x) - logE(x) - 1\right] \tag{10}$$

$$C_t(x,t) = C_t(x,t-1) + \frac{K_{trans'}(x)}{[logE(x)]^2}F(x,t) \tag{11}$$

where $\triangle t$ is the scan interval, $K^{trans}$ is the transfer constant which characterizes the diffusive transport of low-molecular weight Gd chelates across the capillary endothelium,[48] $V_e$ is the fractional volume of the extravascular extracellular space (EES). $C_t(x,t)$ depends on $C_p(t)$, the concentration of contrast agent in the blood plasma (i.e., the AIF), as well as PK parameters, such as $K^{trans}(x)$ and $V_e(x)$. Note that the population AIF, $C_p$, varies over time, whereas PK parameters $K^{trans}$ and $V_e$ depend on different tissue properties and hence vary spatially and anatomically, but do not change over time. The correlation between $C_t(x,t)$ and $C_t(x,t-1)$ is shown in Equation (11), where piecewise linear representation [30] is used to model AIF in Tofts model.[1]

Since DCE-MRI time series is temporally correlated, transformations like the Discrete Cosine Transform (DCT) concentrates its energy in lower frequencies (a few transformed coefficients), as shown in Figure 3a. TSCF makes use of this sparsity in the spectral domain.

*2.2.2 Temporo-Spatial Collaborative Filtering (TSCF)*

The proposed temporo-spatial filtering is a type of transform-domain denoising technique where the true signal intensity can be approximated by a linear combination of few basis elements. This involves three basic steps: 1) Gather, 2) Noise attenuation with collaborative filtering and 3) Reduce. The overall algorithm for TSCF is summarized in Algorithm 1.

---

**Algorithm 1** Temporal Spatial Collaborative Filtering (TSCF)

---

**Input:** A variance stabilized noisy image $\mathbf{Z}$, and the image's signal domain $\mathbf{X} \subset \mathbb{Z}^d$, where $d \in \{2, 3\}$.
**Output:** A denoised image $\hat{\mathbf{Y}}$, which is an estimation of the noiseless image $\mathbf{Y}$.

1: **while** stopping criterion not met **do**
2:    Transform each voxel's DCE-MRI time series (*temporal transform*), $z(x, :)$, into spectral domain, denoted as $\mathscr{T}_t(z(x, :))$.
3:    **for** each reference voxel $x_r \in \mathbf{X}$ **do**
4:      *Gather:*
5:        Calculate spatial distance with respect to the reference voxel, $x_r$, with Equation (12).
6:        Calculate temporal similarity with respect to the reference voxel in spectral domain with Equation (13).
7:        Form a cluster, $P_{x_r}^z$, for the reference voxel $x_r$ based on Equation (14). $C_{x_r}^z$ is the corresponding cluster data in spectral domain, as shown in Equation (16).
8:      *Noise Attenuation with Collaborative Filtering:*
9:        Apply *spatial transform* on each cluster to leverage spatial similarity across different voxels within a cluster, denoted as $\mathscr{T}_s(C_{x_r}^z)$.
10:       Perform empirical Wiener filtering on the temporally and spatially transformed 2-D matrix in spectral domain, with Equation (18) to Equation (20)
11:       Transform filtered signal back into original signal domain with Equation (21). An estimation on the noiseless signal within this cluster is represented by $\hat{G}_{x_r}^z$.
12:    **end for**
13:    *Reduce:*
14:      **for** each voxel $x_i \in \mathbf{X}$ **do**
15:        Find the set of clusters this voxel contributed to, $J_{x_i}$, by Equation (22).
16:        Compute the final estimation for this voxel, $\hat{y}(x_i, :)$, with estimations from different clusters by Equation (23)
17:      **end for**
18: **end while**

---

The following subsections present the sub-steps of our proposed TSCF algorithm in detail.

**a) Gather:** This step involves finding similar voxels in the 2-D (or 3-D) signal domain and combining them across the entire time-series of DCE-MRI input data into collections that we call *clusters*. Hence the gather operation exploits potential similarity (correlation, affinity, etc) in both temporal and spatial domain of DCE-MRI data. The opportunity for such voxel similarity across the image sequence arises because of mainly three reasons: 1) DCE-MRI has a higher temporal resolution compared to the diffusion rate of the contrast agent (CA). Hence, the per-voxel intensity change due to CA diffusion in the entire DCE-MRI image sequence renders the voxels in the same location across different time points correlated to each other, as shown in Equation (7) and Equation (11); 2) Due to anatomical spatial similarity, the pre-contrast voxel intensities of voxels with same or similar tissue types are very close to each other; 3) Finally the anatomical similarity also leads to similarity in pharmacokinetic tissue properties that in turn result in similar CA concentration in the tissue.

The spatial distance between two voxels $x_i$ and $x_j$ are measured by squared Euclidean distance of their coordinates as shown in Equation (12). Let $z(x,:)$ denote the noisy DCE-MRI time series of the voxel at the coordinate $x \in \mathbf{X}$. As shown in Figure 4, in the *Gather* step, the transformation is first applied to each time series $z(x,:)$. Then the similarity between any pair of voxels, $s(x_i, x_j)$, is measured by the squared Euclidean distance between transformed time series in spectral domain as shown in Equation (13), where $\mathscr{T}_t(\cdot)$ is the temporal transformation on the time series of a voxel.

$$d(x_i, x_j) = \|x_i - x_j\|_2^2 \tag{12}$$

$$s(x_i, x_j) = \|\mathscr{T}_t(z(x_i,:)) - \mathscr{T}_t(z(x_j,:))\|_2^2 \tag{13}$$

TSCF loops over all voxels, giving each voxel a chance to become the reference voxel. For each reference voxel $x_r$, its similarity to any other voxel is sorted, and a cluster of similar voxels are formed for this reference voxel based on both the similarity metric and the spatial distance.

As shown in Equation (14), we define the cluster formed using the reference voxel $x_r$ as $P_{x_r}^z$. Since each element in this cluster is already a vector, we organize the voxels in this cluster row by row. Therefore, no matter whether the original noisy signal is in 2-D or 3-D signal space, the cluster of each reference voxel forms a 2-D matrix $G_{x_r}^z$ in the original signal domain as shown in Equation (15), and forms a 2-D matrix $C_{x_r}^z$ in spectral domain as shown in Equation (16).

$$P_{x_r}^z = \{x_i \mid x_i \in \mathbf{X} \wedge s(x_r, x_i) \leqslant \zeta_{\text{sim}} \wedge d(x_r, x_i) \leqslant \tau_{\text{dist}}\} \tag{14}$$

$$G_{x_r}^z = z\left(P_{x_r}^z, :\right) \tag{15}$$

$$C_{x_r}^z = \mathscr{T}_t\left(z\left(P_{x_r}^z, :\right)\right) \tag{16}$$

The significance of the gather operation is to enable collaborative filtering of voxels in each cluster by exploiting the enhanced sparsity in the spectral domain introduced by intelligently incorporating voxel similarity information across both the temporal and spatial domain.

**b) Noise Attenuation with Collaborative Filtering:** Given a cluster of similar voxels, *collaborative filtering* of the cluster produces denoised estimates of the voxels. This has 3 sub-steps 1) Cluster transformation, 2) Noise attenuation with empirical Wiener filtering[49] and 3) Inverse cluster transformation.

1) Cluster transformation: For each cluster formed in the *Gather* step, we firstly apply spatial transformation over voxels in the cluster($\mathscr{T}_s\left(\cdot\right)$), as shown in Equation (17). This operation extracts the spatial similarity and gets $S_{x_r}^z$, which is a 2-D matrix in spectral domain. Because of spatial similarity, $S_{x_r}^z$ is sparse.

2) Empirical Wiener filtering: We use empirical Wiener filtering to attenuate the noise.

$$S_{x_r}^z = \mathscr{T}_s\left(C_{x_r}^z\right) \tag{17}$$

$$H_{x_r}^z(u,v) = \begin{cases} S_{x_r}^z(u,v) & \text{if } \left|S_{x_r}^z(u,v)\right| \geqslant \tau \\ 0 & \text{otherwise.} \end{cases} \tag{18}$$

$$W_{x_r}^z(u,:) = \frac{\|H_{x_r}^z(u,:)\|_2^2}{\|H_{x_r}^z\|_2^2} \tag{19}$$

$$U_{x_r}^z = W_{x_r}^z S_{x_r}^z \tag{20}$$

where $u$ and $v$ are the transform frequencies in 2-D spectral domain. $\tau$ is the threshold for attenuating noise and calculating weights of Wiener filter. Although the weights $W_{x_r}^z$ are calculated on thresholded 2-D spectrum signal $H_{x_r}^z$, the Wiener filtering itself is applied to the original spectrum signal $S_{x_r}^z$ as shown in Equation (20).

3) Inverse cluster transformation: As shown in Equation (21), after Wiener filtering, the 2-D spectrum data is transformed first back into spatial domain ($\mathscr{T}_s^{-1}(\cdot)$), and then into the time domain($\mathscr{T}_t^{-1}(\cdot)$). After these two steps of inverse transformation, we get $\hat{G}_{x_r}^z$, the denoised estimation of $G_{x_r}^z$, in the original signal domain, as shown in Equation (21).

$$\hat{G}_{x_r}^z = \mathscr{T}_t^{-1}\left(\mathscr{T}_s^{-1}\left(U_{x_r}^z\right)\right) \tag{21}$$

**c) Reduce:** In the *Gather* and *Noise Attenuation* steps, each voxel can belong to several different clusters. As shown in 5, the light green voxels belong to both *Cluster A* with reference voxel A, and *Cluster B* with reference voxel B, because of their similarity to voxel A and voxel B in spectral domain. In the *Reduce* step, the final estimation of the DCE-MRI time series is the average of estimations from each cluster it contributed to, as shown in Equation (23).

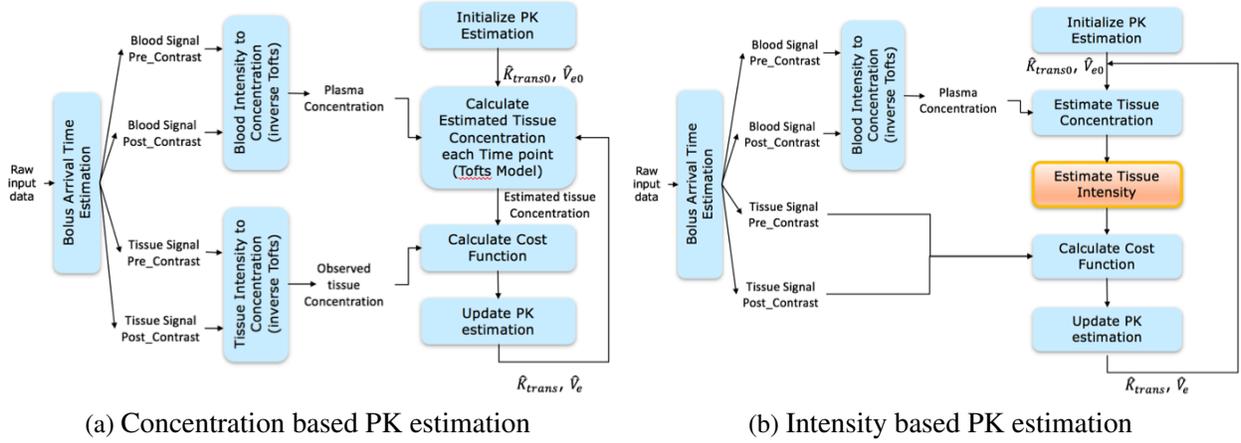

(a) Concentration based PK estimation  (b) Intensity based PK estimation

Fig 6: Different types of PK estimation pipeline

$$J_{x_i} = \{x_j \mid x_i \in P^z_{x_j} \wedge x_i \in \mathbf{X} \wedge x_j \in \mathbf{X}\} \quad (22)$$

$$\hat{y}(x_i,:) = \frac{\sum_{x_j \in J_{x_i}} \left(\hat{G}^z_{x_j}(x_i,:)\right)}{|J_{x_i}|} \quad (23)$$

where $J_{x_i}$ defines a set of reference voxels which $x_i$ clustered with, and $|J_{x_i}|$ is the cardinality of set $J_{x_i}$.

The sequence of *Gather*, *Noise Attenuation*, and *Reduce* forms one iteration of TSCF. This sequence can be configured to run iteratively until the standard deviation of denoised data no longer changes statistically.

The overall signal restoration stage can be summarized in Equation (24).

$$\hat{\mathbf{y}} = \text{IVST}\left(\text{TSCF}\left(\text{VST}\left(\mathbf{z}, \sigma_g\right), \sigma_v\right), \sigma_g\right) \quad (24)$$

*2.3 Pharmacokinetic (PK) Parameter Estimation*

This stage is responsible for accurate and efficient estimation of PK parameters from denoised DCE-MRI data from the *signal restoration* stage.

*2.3.1 Theory*

The change in DCE-MRI intensity due to CA injection is modeled in Equation (2). Given the Gd concentration as a function of time, PK analysis models how the CA distributes in the body, and its dependency on characteristics of the tumor/tissue biology. One well-accepted PK model is the Tofts model (TM),[1] which assumes the diffusible tracer diffuses into the tissue's extracellular space at a rate that is proportional to the instantaneous concentration difference between the plasma and the tissue distribution space, as shown in Equation (25):

$$C_t(t) = K^{trans} \int_0^t C_p(\tau) exp(\frac{-K^{trans}(t-\tau)}{v_e}) d\tau \qquad (25)$$

where $C_t(t)$ is the concentration of CA in the tissue at time t, and $C_p$ is the blood plasma concentration of CA.

The objective of PK analysis is to estimate important PK parameters, such as $K^{trans}$ and $V_e$, from the DCE-MRI tissue intensity signal $\hat{y}$.

*2.3.2 PK Estimation Pipeline*

As shown in Figure 6a, the PK estimation pipeline involves three sub-steps : 1) Bolus Arrival Time (BAT) estimation, 2) blood/tissue concentration calculation and 3) PK parameter estimation. As the output of one sub-step is used in the next, the accuracy in each of these sub-steps is crucial for the overall precision and reproducibility of the PK parameter estimation.

**a) BAT estimation:** Previous studies demonstrate that alignment of the BAT of the AIF and tissue residual function (TRF) can have a large effect on the accuracy of estimation of PK parameters.[50–52] As shown in Figure 1a, with the injection of the CA, the change in its concentration in blood/tissue with time is reflected in the DCE-MRI time series. The objective of BAT estimation is to identify the rising edge in this time varying CA concentration to align the BAT of the AIF and TRF. The fundamental challenge in accurate BAT estimation is that when an image scan is corrupted with noise, it becomes more difficult to identify this rising edge (as shown in Figure 1b) when compared to the case of noiseless signal (as shown in Figure 1a). Hence image denoising is an effective way to improve the accuracy of the BAT estimation.

We used an in-house implementation of the gradient based method[52] to estimate BAT. The only difference is that we perform BAT estimation on intensity values rather than concentration values. This is because the computation of concentration values from intensity values requires correct pre-contrast values, as shown in Equation (26). But the clear separation of pre- and post-contrast data is unknown before BAT estimation. Conversion from intensity to concentration is not guaranteed to be accurate before BAT estimation, so we chose to estimate BAT directly from intensity values, without conversion. Precise BAT estimation is fundamental in accurately separating pre- and post-contrast AIF and tissue intensity, which in turn is important for accurate model fitting to recover the PK parameters.

**b) Compute Tissue/Blood Concentration from Tissue/Blood Intensity:**

Based on Equations (2),(4),(5), and (6), we can compute tissue concentration from tissue intensity as shown in Equations (26) to (28), where $y(x,0)$ is the pre-contrast tissue intensity.

$$\frac{y(x,t)}{y(x,0)} = \frac{[1 - Ae^{D*C_t(x,t)}](1 - B)}{[1 - Be^{D*C_t(x,t)}](1 - A)} \tag{26}$$

$$\gamma(x,t) = \frac{y(x,t)(1 - A)}{y(x,0)(1 - B)} \tag{27}$$

$$C_t(x,t) = \frac{1}{D}\log\left[\frac{1 - \gamma(x,t)}{A - B\gamma(x,t)}\right] \tag{28}$$

If available AIF information is in blood intensity format, we can calculate blood concentration $C_b(t)$ from blood intensity by using $T_1$ of blood in Equation (28), and then compute plasma concentration $C_p(t)$ with hematocrit Hct and $C_b(t)$.

**c) PK Parameter Estimation:** This stage poses the problem of PK parameter estimation as an optimization problem to find the best curve fit of the model using the concentration data of both blood (AIF) and tissue. In this paper, we use constrained nonlinear optimization algorithm L-BFGS[53] to estimate $K^{trans}$ and $V_e$.

As shown in figure 6a, we first randomly initialize the tissue parameter maps ($K^{trans}$ and $V_e$) and then using Tofts model[1] and piecewise linear representation of AIF,[30] calculate the estimated tissue concentration for each time point with Equation (8) to Equation (11). We compare this tissue concentration estimation with true concentration information (i.e., calculate the cost function), and generate new $K^{trans}$ and $V_e$ estimates based on the gradient of the cost function. We continue until the concentration values calculated from estimated $K^{trans}$ and $V_e$ are similar to true data within a user-defined threshold.

There are a couple of steps in the PK parameter estimation that require particular attention for accuracy and computational efficiency. First, the concentration estimation using the Tofts model (Equation (25)) involves an integration step to be computed for each time point in the DCE-MRI sequence and to be repeated for every iteration of the optimization loop. The piecewise linear representation of AIF and recursive calculation of concentration values proposed by Horsfield et al.[30] is much faster than directly performing numerical integration on $C_p$.

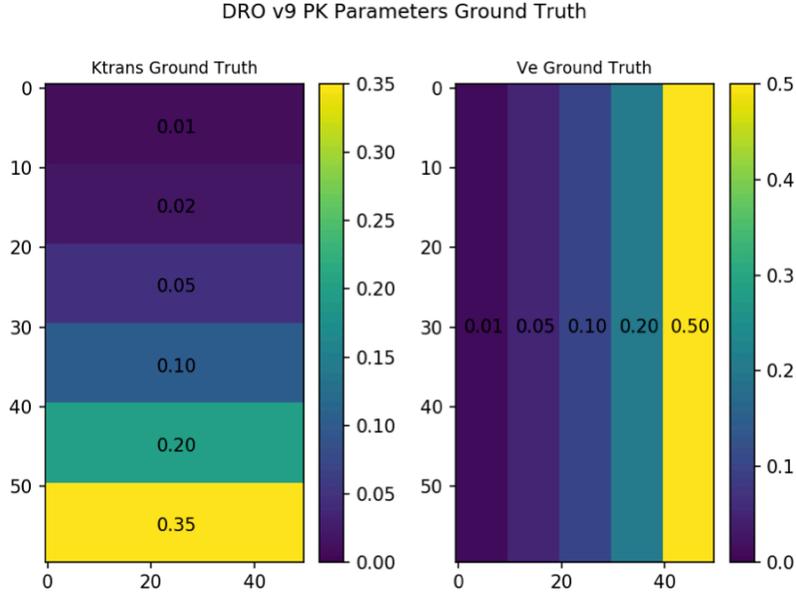

Fig 7: DRO v9 Parameter Ground Truth.

Second, when we use concentration based PK estimation, as shown in Figure 6a, before the main parameter estimation loop, we need to convert the tissue intensity to tissue concentration values once using Tofts model (Equation (26) to (28)). An alternative method is to use intensity based PK estimation, as shown in Figure 6b, where we can save the calculation of concentration from intensity values once before the main PK estimation loop. However, within the main PK estimation iteration, whenever a new $K^{trans}$ and new $V_e$ are estimated, we need to compute tissue intensity from tissue concentration, as highlighted in the orange box. Since this calculation happens in each iteration of the main PK estimation loop, intensity based PK estimation takes longer than concentration based PK estimation. Therefore, we use concentration based PK estimation in our framework.

## 3 Results

*3.1 Data*

*3.1.1 Simulated Digital reference object (DRO) DCE-MRI*

DRO DCE-MRI phantoms with known PK parameter values provide an effective way to validate and compare different DCE-MRI data analyses. The Quantitative Imaging Biomarker Alliance (QIBA) and Duke University have generated a series of DRO DCE-MRI phantoms.[45] We use the `DCE-MRI DRO version 9` for the denoising study in this paper.

In the dataset `DCE-MRI DRO version 9`, noisy data was modeled using Equation (29) (see Appendix). The DRO phantom we use, `QIBA_v9_Tofts_S0_500_6s_0s_sigma_5`, has a sampling interval of 6 seconds, timing offset of 0 seconds, and standard deviation of 5 for the noise. The true $K^{trans}$ and $V_e$ 2D maps of `DCE-MRI DRO version_9` are illustrated in Figure 7. Six $K^{trans}$ values $\{0.01, 0.02, 0.05, 0.1, 0.2, 0.35\}$ and five $V_e$ values $\{0.01, 0.05, 0.1, 0.2, 0.5\}$ vary along the $y-$ and $x-$axis respectively. There are 30 combinations of $K^{trans}$ and $V_e$ in total; each combination occupies $10 \times 10$ pixels. The AIF ROI is in the bottom $50 \times 10$ pixels. The DRO simulated 360 seconds of imaging, with injection of CA occurring at 60 seconds. The MR parameters used in the generation of the DRO are: flip angle $30°$, repetition time 5 ms, a pre-contrast $T_1$ ($T_{10}$) of 1000 ms in tissue and 1440 ms in blood, a blood hematocrit of 45%, and the relaxivity of the gadolinium CA at 1.5 T was assumed to be 0.0045 $\text{mmol}^{-1}\text{ms}^{-1}$.

*3.1.2 Breast cancer DCE-MRI data*

With consent from patients and approval from the local Institutional Review Board (IRB), Oregon Health and Science University (OHSU) as an individual Quantitative Imaging Network (QIN) cen-ter acquired the `TCIA QIN Breast DCE-MRI` dataset from ten patients with locally advanced breast cancer who underwent preoperative neoadjuvant chemotherapy (NACT).[31,54] For each patient, four DCE-MRI sessions were acquired during the treatment course: pre-NACT (visit 1, V1), after the first cycle of NACT (visit 2, V2), at a NACT midpoint (usually after the third NACT cycle; visit 3), and after the completion of NACT (visit 4). After all four DCE-MRI sessions, three patients were identified by doctors as with pathologic complete response (pCR), the rest seven patients are non-pCRs (which includes both pathologic nonresponse and pathologic partial response).

The DCE-MRI images were acquired using a Siemens 3T system with a body coil transmitter and a four-channel bilateral phased array breast coil receiver. According to Huang et al.,[54] DCE-MRI acquisition parameters included $10°$ flip angle, 2.9/6.2 ms echo time/repetition time (TE/TR), a parallel imaging acceleration factor of two, 30 to 34 cm field of view (FOV), $320 \times 320$ in-plane matrix size, and 1.4 mm slice thickness. The total acquisition time was $\sim 10$ minutes for 32 to 34 image volume sets of 112 to 120 slices, each with 18 to 20 seconds temporal resolution. The time of injection of CA (HP-DO3A) [ProHance] IV (0.1 mmol/kg at 2 ml/s) was set to start following acquisition of two baseline image volumes using a programmable power injector and followed by a 20 ml saline flush. A population-averaged AIF was provided, and it was generated by averaging individually measured AIF from an axillary artery in a previous breast DCE-MRI study.[55] The population AIF has a 10x higher resolution compared to the DCE-MRI data. The `TCIA QIN Breast DCE-MRI` dataset are available in the TCIA[56] QIN breast DCE-MRI collection. In this paper, we use publicly available `TCIA QIN Breast DCE-MRI` V1 and V2 data to perform early clinical prediction using therapy response.

## 3.2 Experimental Method and Evaluation

### 3.2.1 Signal Restoration in DRO v9 Rician noisy data

For `DCE-MRI DRO version_9` data, both data and AIF intensities are corrupted with Rician noise. For each scan, we have $50 \times 10$ noise corrupted samples of the same AIF intensity value. Using our PK parameter estimation framework, discussed in detail in previous sections, we estimate the "true" signal which is briefly discussed as follows. First we estimate the noise statistics of the pre-contrast AIF intensity values e.g., white noise sigma ($\hat{\sigma}_g$) and the `AIF_X0` (noiseless pre-contrast AIF) using dual-noise estimation. For post-contrast noisy AIF data, using the estimated $\hat{\sigma}_g$, and the observed mean and standard deviation of the post-contrast AIF intensity, we estimate the true value of the AIF data. We then estimate the `data_X0` (noiseless pre-contrast image data) based on dual-noise estimation from pre-contrast data intensity values. Finally, using our novel TSCF algorithm, we denoise and estimate noiseless signal for post-contrast image data.

- *PK estimation results on DRO v9 data*

To evaluate the efficacy of the proposed TSCF algorithm on signal restoration and hence PK parameter estimation, we compare its performance with the state-of-the-art denoising method, BM4D, commonly used Gaussian spatial smoothing, and the baseline method of parameter estimation without denoising. In the evaluation process we the estimated $K^{trans}$ and $V_e$ values from denoising and data fitting from different algorithms were compared with the ground truth $K^{trans}$ and $V_e$ values that were used to construct the DRO data. We used two performance metrics, namely normalized root mean square error (NRMSE) and mean structural similarity (MMSIM) index which are briefly described below.

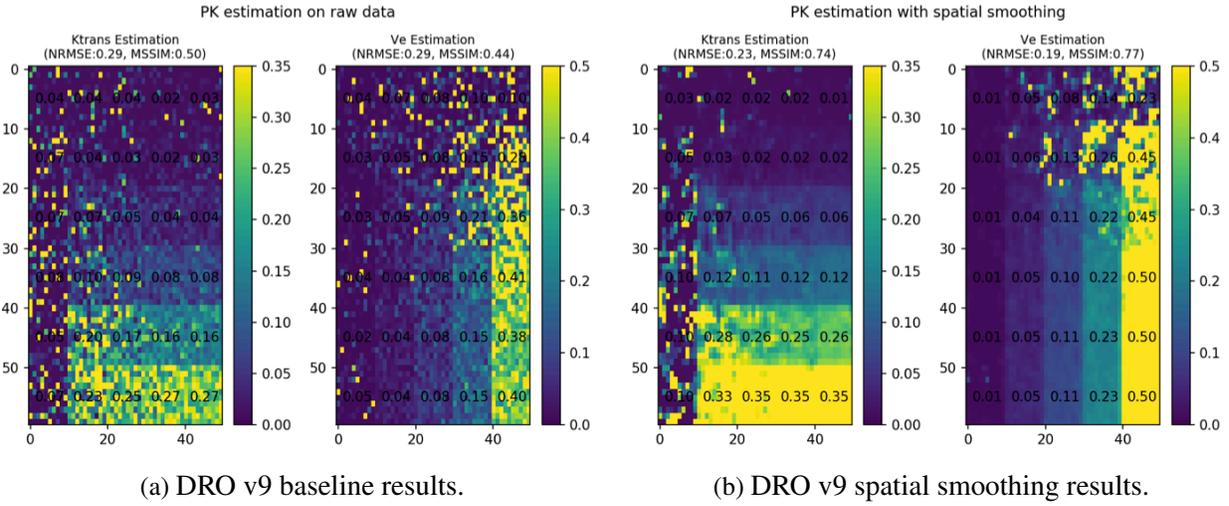

(a) DRO v9 baseline results.

(b) DRO v9 spatial smoothing results.

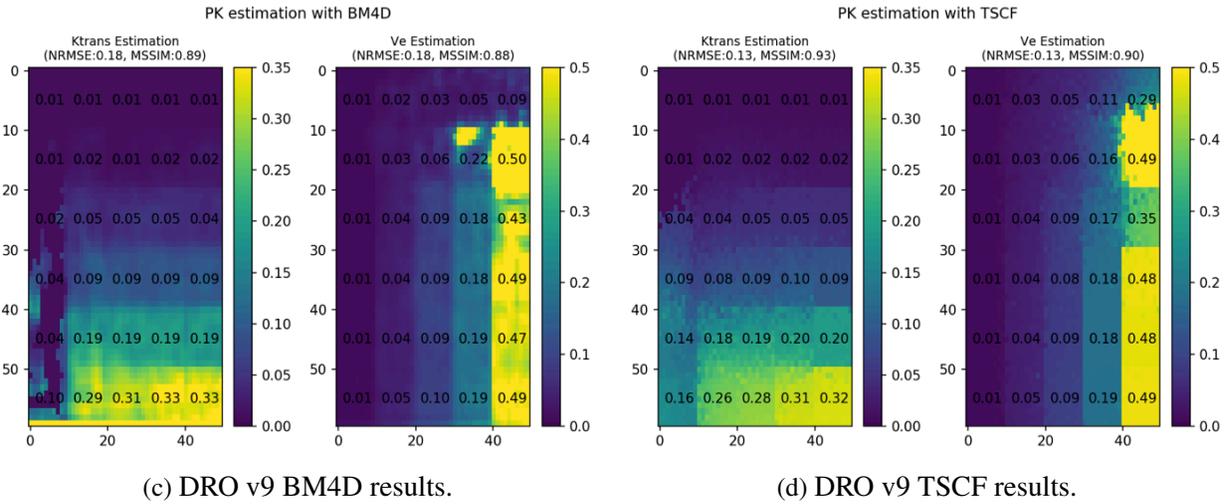

(c) DRO v9 BM4D results.

(d) DRO v9 TSCF results.

Fig 8: DRO v9 results

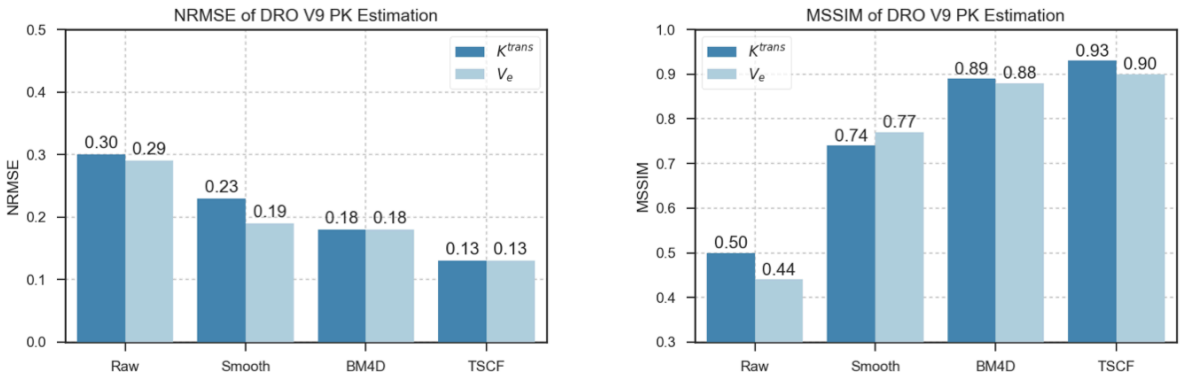

(a) DRO v9 NRMSE results.

(b) DRO v9 MSSIM results.

Fig 9: DRO v9 NRMSE and MSSIM for $K^{trans}$ and $V_e$.

Suppose there are $n$ number of voxels, the true PK value is $\mathbf{x}$, and the estimated PK value is $\hat{\mathbf{x}}$, the NRMSE is defined as $\frac{\sqrt{\frac{\sum_{i=1}^{n}(\hat{x_i}-x_i)^2}{n}}}{\mathbf{x}_{max} - \mathbf{x}_{min}}$. The expected ground truth 2D $K^{trans}$ and $V_e$ map is shown in Figure 7 and the estimated $K^{trans}$ and $V_e$ maps obtained using different denoising algorithms and parameter estimation methods are shown in Figure 8. By visually inspecting Figures 8, we can very easily observe that parameter estimation using our TSCF algorithm (Figure 8d) is much more closer to the ground truth PK parameter in comparison with all other methods. The BM4D method (Figure 8c) visually comes close to our method but not as accurate as TSCF which we will quantify shortly and the baseline method without any denoising technique performs the worst of all(Figure 8a). One of the most commonly used denoising techniques popular for it's simplicity, Gaussian spatial smoothing performs in between (Figure 8b).

In order to quantify the performance we also plot the corresponding NRMSE and MMSIM numbers for each of these methods in Figure 9. We can observe from Figure 9a that the lowest NRMSE (hence the highest accuracy in PK parameter estimation) on the raw DRO v9 data is achieved by our TSCF methods with NRMSE of 0.13 for both $K^{trans}$ (in min$^{-1}$) and $V_e$ which is a substantial improvement of 28% over the state-of-the-art BM4D filtering which has a NRMSE of 0.18 for both $K^{trans}$ and $V_e$.

Several factors contributed towards the better performance of our TSCF algorithm compared to BM4D on DCE-MRI data. First, TSCF makes full use of the information available in the entire dataset by incorporating both spatial and temporal information from the DCE-MRI time series, whereas in BM4D, 3-D images are chopped into small cubes, inside a similarity group, only a small fraction of every voxel's time series is analyzed. Hence, BM4D only utilizes local information, whereas TSCF takes advantage of complete time series information. Secondly, in TSCF, voxel similarity is measured after performing per-voxel time series transformation. This means voxel similarity is measured in spectral domain (Figure 3b, majority energy locates at low frequency components), whereas in BM4D cube similarity is measured in the raw signal domain, which can result in significant inaccuracy in voxel clustering. As shown in Figure 1b, noise corrupted DCE-MRI can completely lose its original CA response pattern and hence it becomes extremely difficult to distinguish voxels with different PK values in the noisy raw signal domain. This makes similarity measure in noisy raw signal domain less accurate. Figure 8a shows that estimating PK parameters by directly applying Tofts model results in the least accurate parameter estimation with the highest NRMSE (0.30 min$^{-1}$ on $K^{trans}$ and 0.29 on $V_e$, Figure 9a) among all the methods. This is because the true signal is corrupted with noise and is used in model fitting without removing the the noise from the signal. Gaussian spatial smoothing (Figure 8b) with *full width at half maximum* (FWHM) of 1.5 improved the NRMSE compared to direct estimation to 0.23 min$^{-1}$ and 0.19 for $K^{trans}$ and $V_e$ (Figure 9a), respectively but falls short compared to more advanced TSCF and BM4D denoising methods. Our TCSF method of PK parameter estimation achieves up to 57% improvement in estimation accuracy.

Besides NRMSE, we also measured MSSIM index[57,58] between estimated and the ground truth PK maps. MSSIM is a perception-based model that considers image degradation as a function of human perceived change in structural information. The key idea is that spatially close voxels have strong inter-dependencies and these dependencies carry important structural information in an image. The higher the MSSIM index value on the estimated PK map, the closer the result is to the ground truth. The advantage of the MSSIM index is that it relates to the human visual system better than mean squared error based methods, and it works in both 2-D images[57] and 3-D images.[58] As shown in Figure 9b, estimating PK values on raw DRO v9 data attains the lowest

Table 1: Parameter settings of TSCF and BM4D

| Parameter | Symbol |
|---|---|
| Image size | $A \times B \times T$ |
| Number voxel | $V$, note $V = A \times B$ |
| Number time points (scans) | $T$ |
| BM4D cube size | $L$ |
| BM4D group size | $M$ |
| BM4D step size | $N_{step}$ |
| BM4D search-cube size | $N_S$ |
| BM4D total num of cubes | $N_c = \left\lceil \frac{A-L}{N_{step}} \right\rceil \left\lceil \frac{B-L}{N_{step}} \right\rceil \left\lceil \frac{T-L}{N_{step}} \right\rceil$ |
| BM4D num of cubes per search-cube | $N_{sc} = \left\lceil \frac{N_S-L}{N_{step}} \right\rceil^3$ |
| TSCF cluster size | $M$ |
| TSCF reduce size | $N$ |
| TSCF number iteration | $K$ |

Table 2: Runtime of TSCF and BM4D

| Image Size | BM4D time (s) | TSCF time (s) | Speedup |
|---|---|---|---|
| $30 \times 30 \times 61$ | 34.5 | 1.4 | 23.9 |
| $40 \times 40 \times 61$ | 73.0 | 3.1 | 23.8 |
| $50 \times 50 \times 61$ | 145.1 | 5.3 | 27.6 |
| $50 \times 60 \times 61$ | 187.8 | 6.5 | 28.9 |

MSSIM index value (0.50 on $K^{trans}$ map and 0.44 on $V_e$ map) amongst all the methods. Spatial smoothing and BM4D further improves this index to up to 0.89 and 0.88 respectively for $K^{trans}$ and $V_e$ respectively. Finally, our TSCF method achieves the highest MSSIM index value with 25.6% and 5% improvement in MSSIM index compared to Spatial smoothing and BM4D, and an over all improvement of 86% over the baseline method. This further reinforces that our proposed TSCF method substantially performs, both quantitatively and qualitatively, better than existing noise canceling and parameter estimation methods on DRO v9 data.

- *Computational Performance Analysis*

To evaluate execution time, we used in house implementation of BM4D and TSCF, both in Python for fair comparison. The experiment is run on a compute node containing two 18-core Intel® Xeon®[1] E5-2699 v3 processors each running at 2.3 GHz and equipped with 256 GB DDR4 memory. We used multi-processing to speedup BM4D using 36 processes. The parameters of BM4D and TSCF are defined in Table 1. For BM4D parameters listed in Table 1, we use the values recommended in the BM4D paper.[44] Although both BM4D and TSCF have time complexity linear with respect to the image size ($\mathcal{O}(VT)$), the constant factors in the complexity function makes the BM4D algorithms significantly more computationally expensive compared to the proposed TSCF algorithm. The sliding window based search method for group formation and the transform in the 4-th dimension renders BM4D ($VT(L^3 M \log L + L^3 M \log M)$) less computationally efficient than TSCF ($VT(\log T + M \log M)$). The execution time of TSCF and BM4D on different sizes of images are presented in Table 2. For each image size, the reported time is the mean runtime of three executions. Overall, TSCF is an order of magnitude faster than BM4D, and the amount of speed up is a function of the image size and other configuration parameters. For `DCE-MRI DRO version_9` data with effective tissue intensity in $50 \times 60 \times 61$ size, TSCF is 28.9X faster than BM4D.

*3.2.2 Signal Restoration in OHSU Breast Cancer Data*

In `TCIA QIN Breast DCE-MRI` data, the contrast agent (CA) was injected after acquisition of two baseline image volumes. However the Bolus Arrival Time (BAT), which is very crucial in accurately aligning AIF and scan data to correctly apply Tofts model, is unknown and varies with different acquisition sessions. Therefore, as discussed in previous sections, data fitting for PK parameter estimation is preceded by BAT estimation. We use a gradient-based method to estimate

---

[1] Intel and Xeon are trademarks of Intel Corporation or its subsidiaries in the U.S. and/or other countries.

BAT for each voxel within a Tumor ROI (TROI), and use the mean BAT across ROI voxels to align tissue and AIF data. Since AIF data has 10x higher resolution compared to tissue data, after calculating BAT on tissue data, we shift AIF data based on 10x of tissue BAT, and then down sample AIF by 10x for final PK estimation.

- *BAT Estimation Results*

When the CA is injected, the tissue concentration and intensity changes due to diffusion of CA and hence we can observe (as shown in 1a) that the CA concentration curve rises, reaches a peak and finally falls with time. As noise can corrupt the CA concentration rising edge of different voxels differently (as shown in Figure 1b), we expect to see higher variance on BAT estimation across voxels within a TROI on noisy data than on denoised data. Hence the variance in the BAT can be used as a metric to estimate the effectiveness of our noise canceling algorithm with a low variance signifying a more accurate BAT estimation that helps in better AIF and data alignment for more accurate model fitting later in the pipeline. As shown in Figure 10, for both visit 1 and visit 2 data, after removing noise with TSCF, the variance of estimated BAT on valid voxels[52] in Tumor ROIs (TROIs) is reduced by 26% and 32%, respectively when compared to without denoising. In other words, TSCF improved the consistency of BAT estimation in TROIs. The key reason behind the variance improvement in visit 2 greater than visit 1 is that the mean of estimated noise standard deviation across subjects, $\overline{\hat{\sigma}_g}$, is larger in visit 2 when compared to visit 1. Hence these results confirm that with our proposed method TSCF, the variance of BAT estimation across voxels within TROIs is substantially reduced in both visit 1 and visit 2 than BAT estimation without noise removal.

- *PK Parameter Estimation Results*

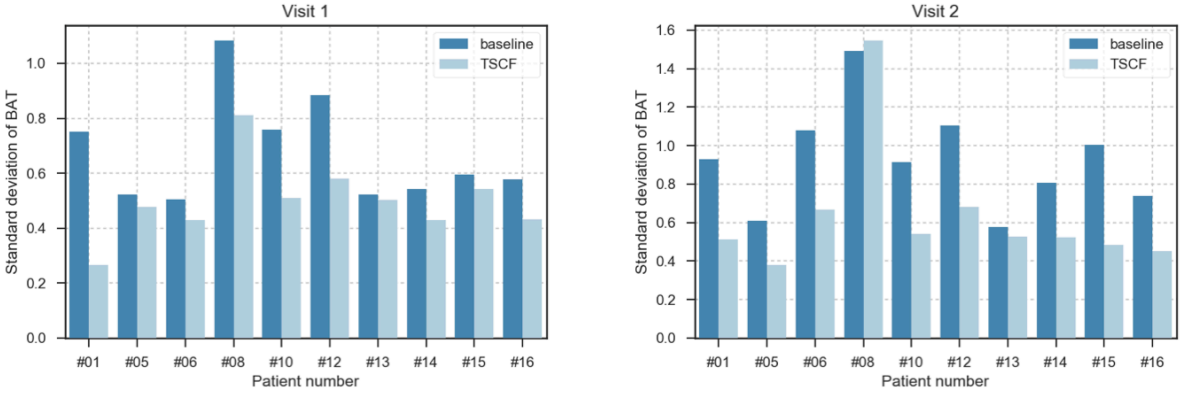

(a) Standard Deviation of BAT on Visit 1.  (b) Standard Deviation of BAT on Visit 2.

Fig 10: Standard deviation of bolus arrival time (BAT) on raw input data and denoised data using TSCF for patient visit 1 and 2.

For the TCIA QIN Breast DCE-MRI dataset, Huang et al.[31] reported the PK estimation results from twelve software tools[31] in their multicenter variations study of DCE-MRI for estimating breast cancer response. Six out of twelve software tools used Tofts model (TM) based PK estimation which is also the model of choice for our proposed PK estimation framework. Figure 11 shows box plot for the estimated $K^{trans}$, $V_e$ and $k_{ep}$ values for V1 and V2 and the corresponding percentage change between V2 and V1 are shown in Figure 12. Although there are no ground truth on $K^{trans}$, $V_e$ and $k_{ep}$ values for this real data, our results are very much in the same range of $K^{trans}$, $V_e$ and $k_{ep}$ values as obtained from different TM based PK estimations in Huang et al.[31] Similar to TM based results reported by Huang et al.,[31] from visit V1 to V2, the mean tumor $K^{trans}$ and $k_{ep}$ value decreases, while the mean tumor $V_e$ value increases in the TCIA QIN Breast DCE-MRI dataset. With TSCF, in most cases ($K^{trans}$ V1, $K^{trans}$ V2, $V_e$ V2 and $k_{ep}$ V2), the variance in the estimated pharmacokinetic parameter values decreases significantly, showcasing the effectiveness of TCSF on real data. We will next explore the advantage of using our framework on end-to-end prediction of therapeutic response.

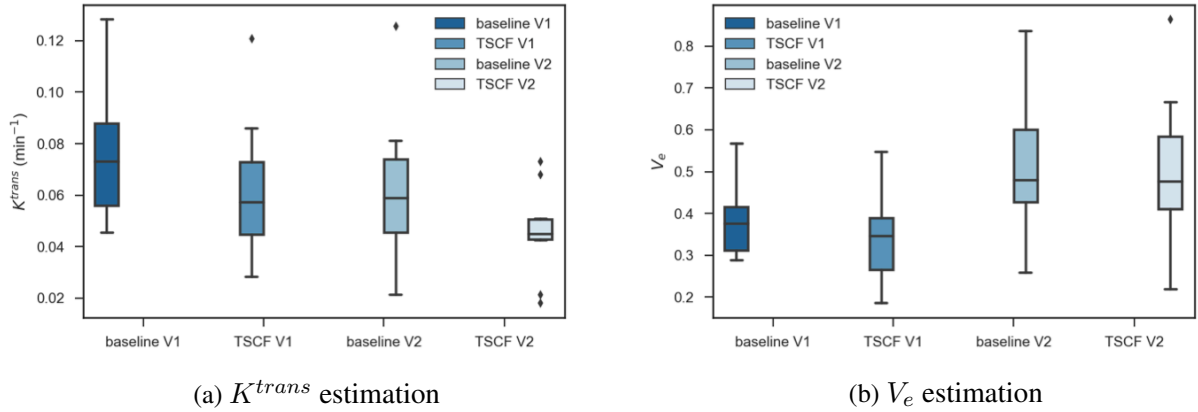

(a) $K^{trans}$ estimation

(b) $V_e$ estimation

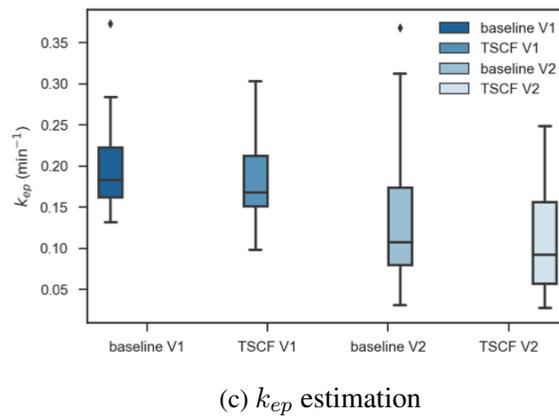

(c) $k_{ep}$ estimation

Fig 11: PK estimation results on TCIA QIN Breast DCE-MRI data

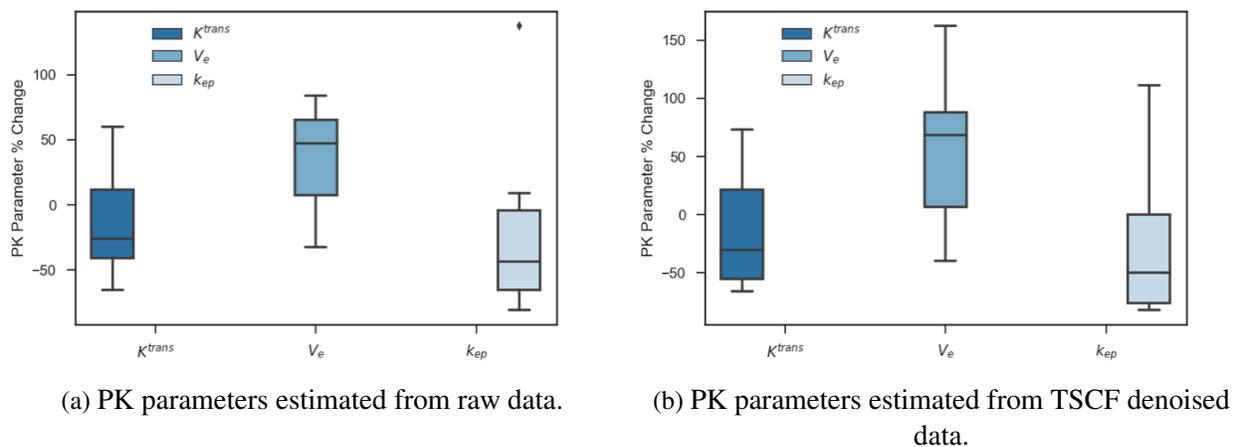

(a) PK parameters estimated from raw data.

(b) PK parameters estimated from TSCF denoised data.

Fig 12: Percentage change in mean tumor $K^{trans}$, $V_e$ and $k_{ep}$ between visit 1 and 2 with raw data and with our TSCF algorithm. The central bar represents the median values.

Table 3: DCE-MRI Parameter univariate logistic regression(ULR) c-Statistic Value for Early Prediction of Pathologic Response

| Parameter | Algorithm | V1 | V2 | %Change(V21) |
|---|---|---|---|---|
| $K^{trans}$ | BAT_TM | 0.524 | 1.0 | 0.952 |
| | TSCF_BAT_TM | 0.619 | 1.0 | 1.0 |
| $V_e$ | BAT_TM | 0.619 | 0.857 | 0.714 |
| | TSCF_BAT_TM | 0.619 | 0.905 | 0.714 |
| $k_{ep}$ | BAT_TM | 0.524 | 1.0 | 0.952 |
| | TSCF_BAT_TM | 0.619 | 1.0 | 1.0 |

- *Prediction of therapeutic response*

Early prediction of therapeutic response in cancer patients can be tremendously useful and possibly life saving. It helps doctors to better plan personalized treatment for different patients, and it helps patients to avoid unnecessary chemotherapy. Similar to the multicenter DCE-MRI analysis challenge,[31] our objective is also to perform early prediction of pCR using only the first two DCE-MRI sessions.

The univariate logistic regression (ULR) $c-$statistic values for early discrimination of pCR and non-pCR are listed in Table 3. For each PK parameter, $K^{trans}$, $V_e$ and $k_{ep}$, we list $c-$statistics based on V1 only, V2 only, as well as its percentage change (V21). *BAT_TM* means BAT estimation and Tofts model based PK estimation are performed on raw data without denoise. *TSCF_BAT_TM* means we first denoise with TSCF, then perform BAT estimation and Tofts model based PK estimation. Our results confirm the observations from[31] that 1) none of the PK parameters at pre-NACT (V1) provide good ($0.8 \leq c < 0.9$) to excellent ($0.9 \leq c \leq 1.0$) discrimination; 2) V2 and V21 provide much better discrimination than V1, and they are able to provide good to excellent discrimination; 3) $K^{trans}$ and $k_{ep}$ provide much better discrimination than $V_e$. More importantly, our results show that for all PK parameters and all cases, V1, V2 and V21, the $c-$statistic values from TSCF is either same or higher than the $c-$statistic values from baseline case without TSCF.

Figure 13 and Figure 14 respectively presents the scatter plots of the percentage changes (V21) on mean tumor $K^{trans}$ and $k_{ep}$, and that for each patient. From Figure 13b, we can observe that by directly using raw data without any noise removal to separate pCRs from non-pCRs show some overlap in the data point and hence isn't clearly separable. On the other hand with TSCF, as shown in Figure 13b, Figure 14c, and Figure 14d, we can completely separate pCRs from non-pCRs using V21 data with a linear decision boundary. To quantify this, from Table 3 we can observe that ULR $c-$statistic for percentage change in both $K^{trans}$ and $k_{ep}$ values achieve ideal value of 1 and an improvement of up to 18% over the prediction using noise corrupted real data.

Figure 15 shows the color tumor $K^{trans}$ maps of non-pCRs (Figure 15a) and pCRs (Figure 15b) at visit V1 and V2 using our proposed TSCF denoising method. The tumor $K^{trans}$ map of each subject is generated for all tumor ROIs on the entire DCE-MRI time series and all $K^{trans}$ maps are overlaid on pre-contrast DCE-MRI images. For each subject at each visit, the $K^{trans}$ map is shown on the slice through the center of the tumor. As shown in the color maps, the $K^{trans}$ values for pCRs in the tumor ROI clearly gets reduced between visit V1 to V2 (for example, red to blue), while for non-pCRs very small variation (both positive and negative) in their $K^{trans}$ values is observed. These observations are consistent with the expected result that as $K^{trans}$ is usually higher in tumor tissue than in normal tissue, and the pCRs completely respond to the therapy, we observe a fall in the $K^{trans}$ values in the tumor ROI between visit V1 and V2. On the other hand, the non-responder non-pCRs have small to no change in their $K^{trans}$ values. The color $k_{ep}$ maps, which have been omitted due to space limitation, also show very similar pattern for both the pCRs and non-pCRs.

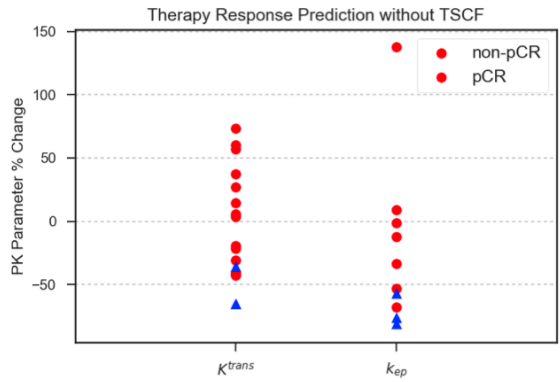
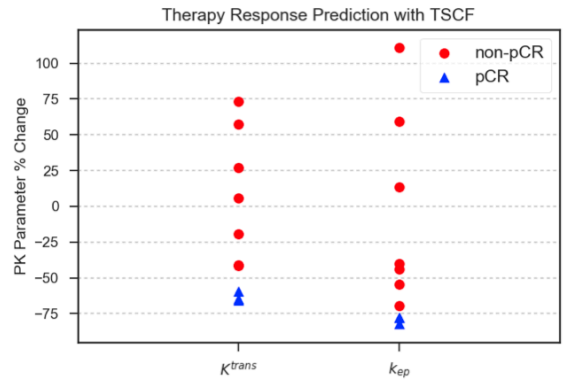

(a) without TSCF

(b) with TSCF

Fig 13: Scatter plots of percentage change in tumor $K^{trans}$ and $k_{ep}$ between visit 1 and 2 for early prediction of pathologic response. Three pCRs are shown as blue triangles and seven non-pCRs are shown as red circle.

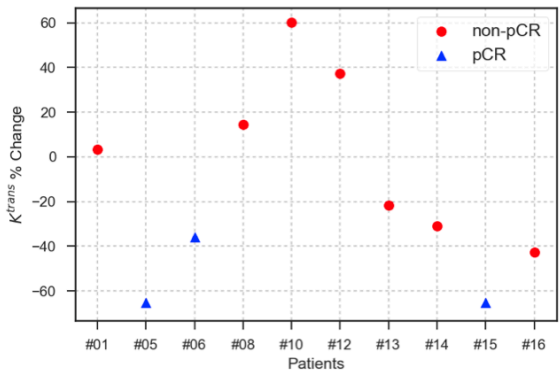
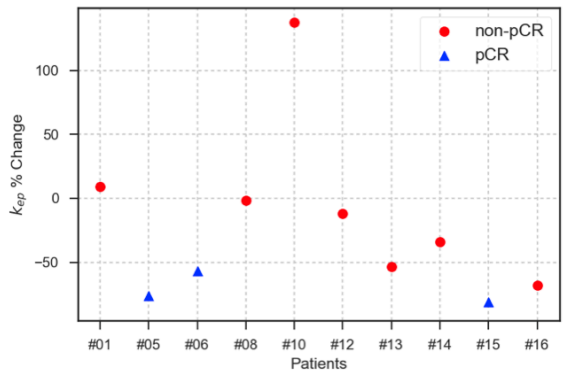

(a) without TSCF

(b) without TSCF

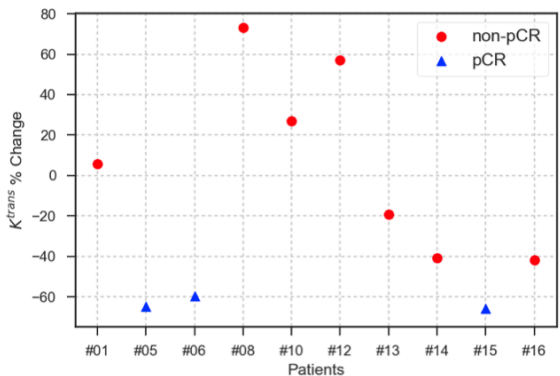
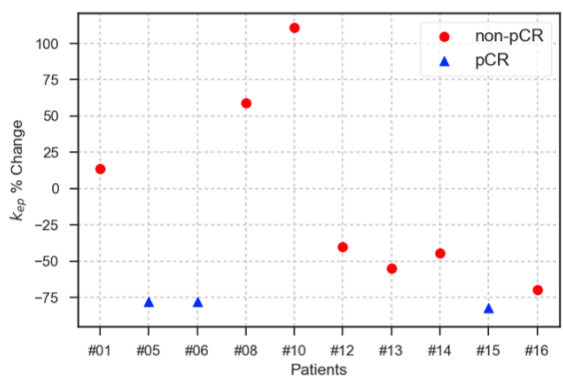

(c) with TSCF

(d) with TSCF

Fig 14: Scatter plots of percentage change in tumor $K^{trans}$ and $k_{ep}$ between visit 1 and 2 for early prediction of pathologic response. Three pCRs are shown as blue triangles and seven non-pCRs are shown as red circle

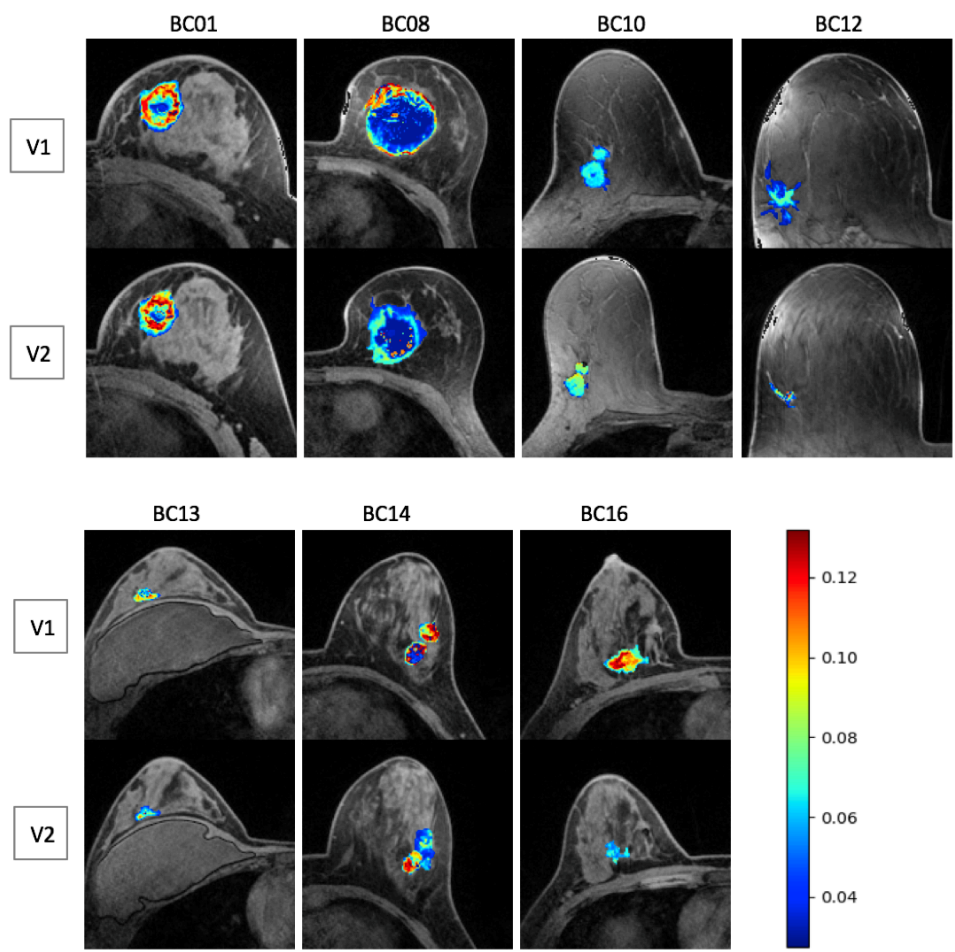

(a) non-pCR

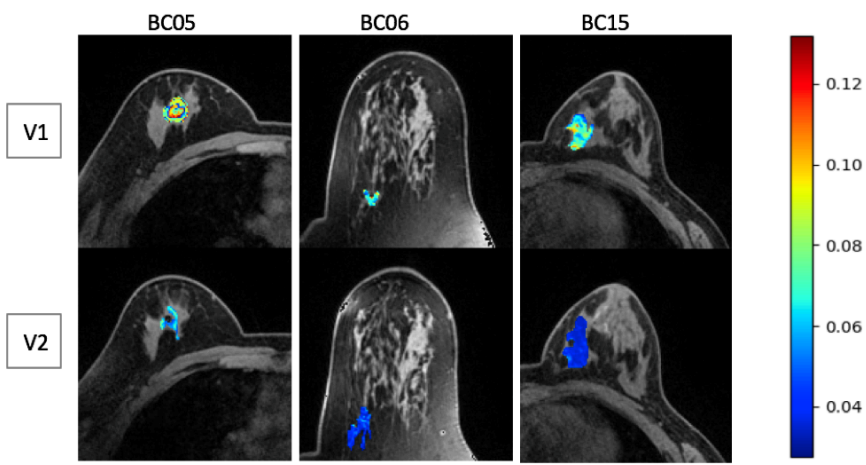

(b) pCR

Fig 15: V1 and V2 single slice tumor $K^{trans}$ parametric map. The color $K^{trans}$ maps are overlaid on pre-contrast DCE images.

## 4 Discussion

TSCF is designed specially for DCE-MRI data. BM3D/BM4D are general method with no specific image type in mind they were designed. Although both TSCF and BM4D utilize spatial similarity, TSCF takes advantage of unique temporal property of DCE-MRI. As shown in Figure 8, Figure 9 and Table 2, TSCF is more accurate and faster than other denoising methods such as BM4D. In DCE-MRI, for each voxel, we capture a series of scans before and after CA injection. The energy of the noiseless version of this time series is primarily linked to low frequency components, as shown in Figure 3a. When the DCE-MRI signal is corrupted with noise, energy also shows up at higher frequencies. As shown in Figure 3, the value of the DC component also changes because of noise. For voxels with similar PK parameter values, lower frequency components are quite similar. Even when they are corrupted with noise, the low frequency components still contain most of the energy. Since the noise of different voxels are independent, the noise triggered high frequency components can show up as different patterns. This property inspired us to do collaborative filtering to attenuate noise on the time series of a group of similar voxels. We consider the problem in both the spatial/voxel and temporal domain. In Particular, we use the transformation of each voxel's time series as the signature of the voxel. In BM4D, noisy observations are directly tested for similarity, and this similarity is used for grouping cubes. A time series of a voxel is analyzed portion by portion in BM4D method, without using the full characteristic of each time series.

Our PK estimation results are consistent with recent observations[59–61] that pharmacokinetic biomarkers have earlier response time than more routine standard of care metrics, such as tumor size changes. As shown in the color tumor $K^{trans}$ maps (Figure 15b), two out of three pCRs have the ROI size increased from V1 to V2 (e.g. for patient BC15 the number of voxels in tumor ROI increased from 3410 to 10577), which might have indicated a radiologist of a non-response to the therapy if only routine standard metrics like tumor ROI size is used for evaluation. Using our TSCF based pharmacokinetic parameter estimation, the percentage change of $K^{trans}$ values from V1 to V2 correctly predicted them as pCRs and hence successfully carrying out a correct early therapy response prediction in the subjects.

Another key thing to note is that in this paper we have proposed a more general denoising paradigm *gather-noise attenuate-reduce* (GNR) and our denoising algorithm TSCF, is an instantiation of the GNR paradigm, which is specifically tailored for DCE-MRI sequences. In GNR we have introduced three primitives: 1) *gather*, responsible for grouping pixels(or voxels) using some similarity metric that helps in advanced noise filtering by taking advantage of the grouping attributes, 2) *noise attenuate*, uses the special attributes of these grouped pixels (e.g. sparsity in spectrum domain) to collaboratively denoise them and finally 3) *reduce* operation to recover the true pixel intensities as a function of the contributions from different denoised groups. Using these three primitives one can implement domain specific denoising algorithms targeted to different applications and image modalities. In sections 2.2.2, we describe in detail TSCF as one specific instantiation of GNR by showing the implementation of these three primitives that suits DCE-MRI sequences and discuss in detail the reason behind the implementation decisions for each on these primitives (e.g. extracting sparsity in *gather* step incorporating both temporal and spatial information in DCE-MRI sequence). BM3D/BM4D can be viewed as an instantiation of GNR that incorporates only spatial information of the image with more exhaustive pixel grouping implementation of *gather* step that might be suitable for applications with stationary or static image modalities. Hence, one can formulate new denoising algorithms for different image modalities by just implementing the GNR primitives.

## 5 Conclusion

DCE-MRI is a minimally invasive imaging technique which can be used for characterizing microvasculature characteristics of normal and cancerous tissues. PK estimation is widely used in DCE-MRI data analysis to extract quantitative parameters relating to vasculature and tissue hemodynamics. The noise in DCE-MRI data has a large effect on the accuracy of PK estimation. In this paper, we proposed a 3-stage denoising and PK parameter estimation pipeline for noise corrupted DCE-MRI data. We proposed a general denoising paradigm, *gather-noise attenuate-reduce* (GNR), for signal restoration, as well as TSCF, a specially designed algorithm to denoise DCE-MRI data. TSCF takes advantage of sparsity brought by temporal correlation in DCE-MRI, as well as anatomical spatial similarity to collaboratively filter noisy DCE-MRI data. We evaluated the proposed framework with both DRO phantom data and `TCIA QIN Breast DCE-MRI` dataset. The proposed TSCF denoising algorithm achieves up to 57% reduction of normalized estimation error, and up to a 86% improvement in the structural similarity of PK parameter estimation than the baseline method without denoising, and an order of magnitude faster than the state-of-the-art denoising method. With TSCF denoising, the ULR $c-$statistic value for early prediction of pathologic response is improved up to 18%. pCRs and non-pCRs in `TCIA QIN Breast DCE-MRI` dataset can be 100% completely separated.

In the future, transformation methods other than DCT will be investigated for TSCF. Also, TSCF will be used as a preprocessing step on noisy DCE-MRI before performing other medical imaging analysis tasks such as image registration/motion estimation. We will also evaluate TSCF for other type of medical images with temporal correlation.

**Disclosures**

Xia Zhu, Dipanjan Sengupta, and Theodore L. Willke are employed by Intel Corporation. The authors have no other conflicts of interest.

*Acknowledgments*

The authors of this paper would like to thank investigators at OHSU, especially Dr. Wei Huang and Dr. Ian Tagge, for providing `TCIA QIN Breast DCE-MRI` dataset and annotations, and for their help in explaining the details of acquisition and meta-data of the `TCIA QIN Breast DCE-MRI` dataset.

**Appendix A:**

*A.1 MR Noise Review*

The MR complex raw data is measured through a quadrature detector resulting in the real and the imaginary quadrature channels. Taking the magnitude of the complex data results in a nonlinear mapping between the noise and the true unknown signal. And hence the additive noise from each quadrature channels is transformed into Rician noise.[62] More formally, the magnitude MR signal $z \sim \mathcal{R}(y, \sigma_g)$, has Rician distribution with parameters $y \geq 0$ and $\sigma_g \geq 0$,

$$z = \sqrt{(y + n_r)^2 + n_i^2} \tag{29}$$

where $y$ is the unknown true signal intensity and $n_r$ and $n_i$ are the noise from the real and complex MR signals with Gaussian distribution of mean zero and standard deviation $\sigma_g$ and the noise variance is assumed to be the same in both the channels.[46,63]

Let say the signal-to-noise ratio (SNR) is defined as $\theta = \frac{y}{\sigma_g}$. According to Koay,[34] the raw image standard deviation $\sigma_r$ is related to $\sigma_g$ as follows:

$$\sigma_r^2 = \xi(\theta)\sigma_g^2 \qquad (30)$$

where $\xi$ is a correction factor which is a function of the SNR $\theta$:

$$\begin{aligned}\xi(\theta) = 2 + \theta^2 - \frac{\pi}{8} \\ \times exp(-\frac{\theta^2}{2})((2+\theta^2)I_0(\frac{\theta^2}{4}) + \theta^2 I_1(\frac{\theta^2}{4}))^2\end{aligned} \qquad (31)$$

where $I_n$ denotes the Modified Bessel function of order $n$, $I_n(x) = \sum \frac{(x/2)^{n+2m}}{m!\Gamma(n+m+1)}$.[64]

Koay[34] further derives a fixed point formula of SNR, that establishes a relationship between true signal SNR ($\theta = \frac{y}{\sigma_g}$) and magnitude SNR ($\frac{\mu_m}{\sigma_m}$):

$$\theta = \sqrt{\xi(\theta)[1 + \frac{\mu_m^2}{\sigma_m^2}] - 2} \qquad (32)$$

where $\mu_m$ is first moment (observed mean) of the magnitude DCE-MRI data, $\sigma_m$ is the standard deviation of the magnitude signal. Finally the fixed point solution of SNR can be calculated using Newton's method of root finding if the mean and the standard deviation of the magnitude signals are available.

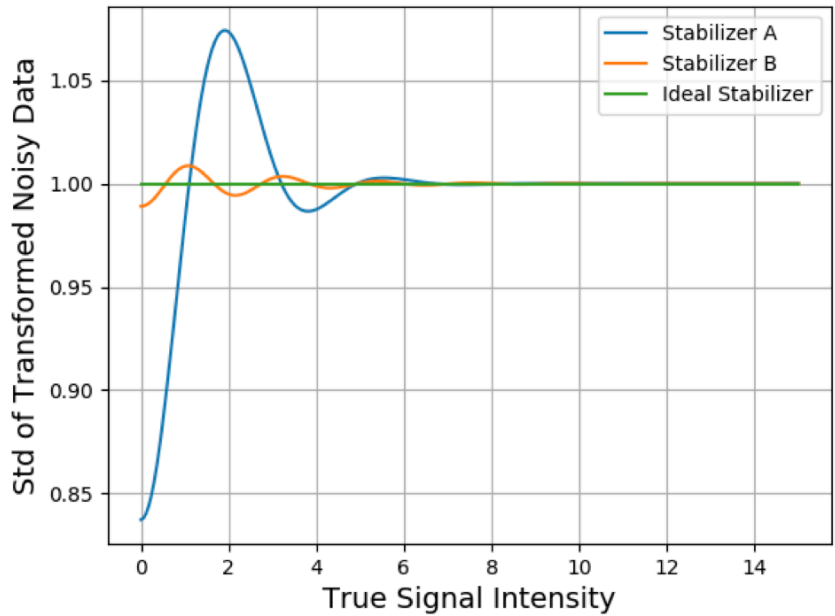

Fig 16: Variance Stabilizer.

*A.2 Variance Stabilizer*

For both VST and IVST step, we use the stabilizer 'B' from,[35] because compared to stabilizer 'A', it approximates ideal stabilizer much better when the ratio of the true signal to the standard deviation of transformed noisy signal is less than 6, as shown in Figure 16.

**Xia Zhu** is a senior research scientist at Intel Corporation. She received her PhD in computer science in 2003. She was a senior research scientist in Phillips Research from 2003 to 2005. She joined in Intel research lab in 2005. Her current research interests include medical image analytics, machine learning, brain decoding and graph analytics.

**Dipanjan Sengupta** is a research scientist at Intel Corporation. He received his Bachelor of Technology in computer science and engineering from Indian Institute of Technology Kharagpur (IIT KGP) and PhD in computer science from Georgia Institute of Technology in 2008 and 2016, respectively. His current research interests include medical imaging, graph theory and machine learning.

**Andrew Beers** is a staff programmer at the Quantitative Tumor Imaging Lab at the Martinos Center. His current research interests include developing open-source deep learning software for neuroimaging problems, tumor and anomaly segmentation in medical images, and interpretability in deep learning for medical applications.

**Kalpathy-Cramer Jayashree** is a researcher working at the intersection of artificial intelligence and medical imaging. She is the co-director of the Quantitative Tumor Imaging at Martinos lab at the Athinoula A. Martinos Center for Biomedical Imaging at Massachusetts General Hospital and a principal investigator affiliated with the Center for Clinical Data Science. Her areas of research include quantitative imaging in cancer, image analysis for retinal imaging, cloud computing, and machine learning. Jayashree has a Bachelor of Technology in electrical engineering from IIT Bombay, a Master of Science in biomedical informatics from Oregon Health & Science University, and a Ph.D. in electrical engineering from Rensselaer Polytechnic Institute.

**Theodore L. Willke** leads a team that researches large-scale machine learning and data mining techniques in Intel Labs. His research interests include parallel and distributed systems, image processing, machine learning, graph analytics, and cognitive neuroscience. He has authored over 40 papers on related topics. He is also a co-principal investigator in a multi-year grand challenge project on real-time brain decoding with the Princeton Neuroscience Institute. Previously, he founded an Intel venture focused on graph analytics for data science that is now an Intel-supported open source project. In 2014, he won Intels highest award for this effort. In 2015, he was appointed to the Science & Technology Advisory Committee of the US Department of Homeland Security. He holds a doctorate in electrical engineering from Columbia University, a masters from the University of Wisconsin-Madison, and a bachelors from the University of Illinois.

# List of Figures

# List of Tables